\documentstyle[preprint,tighten,version2,aps]{revtex}
{\makeatletter%
\gdef\labeleqs#1{{%
\edef\@currentlabel{%
\ifappendixon\appletter\fi
\ifsecnumbers\ifnum\c@secnum>0
\arabic{secnum}.\fi\fi\arabic{equation}}%
\label{#1}%
}}%
}
\begin{document}
\draft
\preprint{IFUP-TH 30/94}
\begin{title}
\hskip-10pt Detecting Dual Superconductivity in the Ground State\\
 of Gauge
Theories - II.\footnote{Partially supported by MURST\\
Contribution to the 27-th International Conference on High Energy
Physics, Glasgow July 1994}
\end{title}
\author{L. Del Debbio, A. Di Giacomo, G. Paffuti, P. Pieri}
\begin{instit}
Dipartimento di Fisica dell'Universit\`a and
I.N.F.N., I-56126 Pisa, Italy
\end{instit}
\begin{abstract}
A monopole creation operator is constructed: its vacuum expectation
value is an order parameter for dual superconductivity in that, if
different from zero it signals spontaneous breaking of the U(1)
symmetry corresponding to monopole charge conservation.
The operator is tested on compact U(1) gauge theory on lattice.
For SU(2) gauge theory it clearly demonstrates that confinement is
produced by dual superconductivity.
\end{abstract}
%
%
\section{Introduction}
A possible mechanism of colour confinement in Quantum Chromo-Dynamics
({\it QCD\/}) is dual superconductivity of the vacuum~\cite{tHO,Man,Nie,Rev}.
According to this scenario, the chromoelectric field is channelled
into Abrisokov flux tubes~\cite{Abri}, in the same way as the ordinary magnetic
field is in superconductors: the word {\it dual} indicates the
interchange of roles between electric and magnetic fields and charges.
The chromoelectric field mediating the force between coloured
particles is squeezed by Meissner effect into flux tubes of constant
energy per unit length, giving rise to the confining linear potential.
These flux tubes behave as strings~\cite{Nie,Nam}.
The existence of strings in hadronic physics is supported by phenomenology
\cite{Ven,Reb}. They have also been visualized by numerical simulations of QCD
on the lattice~\cite{dgmo,r10}. Some evidence in favour of dual
superconductivity of the vacuum has been produced by Montecarlo
simulations~\cite{r11}.

A clear cut test of the mechanism would be the detection of monopole
condensation in the ground
state, analogous to the condensation of Cooper pairs in the ground
state of an ordinary superconductor. Condensation implies that the
vacuum is a superposition of states with different charge, which, in
turn, is nothing but a spontaneous breaking of the $U(1)$ symmetry
related to charge conservation~\cite{Wein}. Such a breaking is
signaled by a non-vanishing vacuum expectation value
({\em vev \/}) of any operator carrying
non-trivial charge, as, e.g. the scalar field of the Landau-Ginzburg
model of superconductivity. That {\em vev} is called a disorder parameter
in the language of statistical mechanics.

In {\it QCD\/} the monopoles which are
expected to condense and generate superconductivity are Dirac
monopoles of a residual $U(1)$ symmetry, which survives after a
suitable gauge fixing, known as abelian projection~\cite{tHO}. An abelian
projection is defined as the gauge transformation which diagonalizes
any operator transforming in the adjoint representation of the gauge
group. The points where two eigenvalues of that operator coincide
correspond to singularities of the gauge transformation and have the
topology of world-lines of point monopoles~\cite{tHO,Schier}.
Such monopoles have been observed on the
lattice~\cite{r16}. Of course the location and the
number of  monopoles do depend on the choice of the operator used to define the
abelian projection: a possibility\cite{tHO} is that
physics is independent of it. The
relevant abelian degrees of freedom
 can also be fixed by a somewhat different
procedure, known as {\em maximal abelian \/}
projection~\cite{Schier}.
Our operator allows to investigate unambigously what abelian projection,
if any, defines the monopoles relevant to confinement.

In this paper,  we will present the construction of the operator,
which is rather general and can be used for any kind of solitons.
We will analyze how the operator works in compact $U(1)$ gauge
theory\cite{nostro}.

For that theory,
there exists a construction of a disorder variable describing monopole
condensation~\cite{FM}, which is rigorous but based on a
particular form of the action (the Villain action). Our operator
coincides with the above one in the Villain case, but can be used with
different forms of the action. This is particularly important in order
to use it for non-abelian theories, where the effective $U(1)$ action
after abelian projection is not known.

For $U(1)$ we get a spectacular signal of monopole condensation\cite{nostro}.
We
then apply the same construction to $SU(2)$ gauge theory: we consider the
monopoles corresponding to the abelian projection in which the Polyakov loop is
diagonal, and we investigate the condensation of monopoles across the
deconfining phase transition. Here again we find a spectacular signal of
monopole condensation. The monopoles corresponding to that
abelian projection do condense in the confined phase: Confinement is produced
by
dual superconductivity.

The next steps of our investigation, which is in progress, will be to study the
behaviour of monopoles defined by different abelian projections, to test
t'Hooft
ideas, and, in particular, the properties of the maximal abelian projection.

In sect. 2 the construction of the monopole creation operator is presented.
Sect.3 contains the results for compact $U(1)$. Sect.4 the results for monopole
defined by abelian projection diagonalising the Polyakov loop in $SU(2)$ gauge
theory. Sect.5 contains a
few concluding remarks.
\section{The Monopole Creation Operator.}
For the sake of definiteness, we shall consider here only the case of
a $U(1)$ monopole. Our procedure can easily be generalized to all kind
of solitons.

Let $B_{\mu}({\bf x}, {\bf y}) = (0, {\bf b}({\bf x}, {\bf y}))$ be
the classical field produced
in the location $\bf x$
by a Dirac monopole at
rest in  ${\bf y}$. We can make any choice for the gauge, e.g. by
putting the string along the positive $z$-axis. Then, defining
${\bf r} = {\bf x} - {\bf y}$ and $r =\|{\bf r}\|$, we have:
\begin{equation}
\label{eq:Dirac}
b_{i}({\bf x}, {\bf y}) = g\: \varepsilon_{3ij} \frac{r_{j}}{r(r-r_3)}
\end{equation}
where $g$ is the charge of the monopole, satisfying the Dirac
quantization condition $e\, g = \frac{n}{2}$.
If $\Pi_{i}({\bf x}, t)$ is the conjugate momentum to $A_{i}({\bf x},
t)$, then the operator
\begin{equation}
\mu({\bf y}, t) = \exp\{i \int d^{3}{\bf x} \; b_{i}({\bf x}, {\bf y}) \,
\Pi_{i}({\bf x}, t)\}
\end{equation}
creates a monopole at the location ${\bf y}$ and time $t$. This can be
immediately seen in the Schr\"odinger representation of the fields,
where we have:
\begin{equation}
\label{eq:trasl}
\mu({\bf y}, 0) |{\bf A}({\bf x}, 0)\rangle =
|{\bf A}({\bf x}, 0) + {\bf b}({\bf x}, {\bf y})\rangle
\end{equation}
Equation~(\ref{eq:trasl}) is a trivial consequence of the canonical
commutation relations between the fields and their conjugate momenta,
and is nothing but the field-theoretic equivalent of the familiar
statement that $e^{i\,pa}$ translates the
coordinate $q$ by $a$.
$\mu({\bf y}, t)$, applied to any field configuration, adds a monopole
to it.
This can be restated in terms of commutation relations as:
\begin{eqnarray}
\left[A_{i}({\bf x}, t), \mu({\bf y}, t)\right] &=&
b_{i}({\bf x},{\bf y}) \mu(\bf{y}, t)\\
\left[\Pi_{i}({\bf x}, t), \mu({\bf y}, t)\right] &=& 0
\end{eqnarray}
which also show that the electric field of the configuration is left
unchanged. It is worthwhile to notice that the specific choice of the
gauge for ${\bf b}$ is irrelevant: what really matters here is
topology, which is independent of it. Our operator $\mu$ is similar to
operators
introduced in the literature by different constructions, in various contexts
\cite{r22}.

\noindent
Now, if the ground state of the theory has a definite monopole number
$N$, then, under a magnetic $U(1)$ rotation, we have:
$U |0\rangle = e^{i \,\varphi \,N} |0\rangle$. Since
$U \mu \, U^{\dag} = e^{i \,\varphi} \mu$,
then
$\langle 0| \mu |0\rangle = \langle 0| \mu |0\rangle e^{i \varphi}$
which implies
$\langle 0| \mu |0\rangle = 0$.
Therefore if
$\langle 0| \mu |0\rangle \neq 0$
then $|0\rangle$ is not $U(1)$ invariant and there is spontaneous
symmetry breaking of $U(1)$.
A translation by a static field ${\bf g}({\bf x})$ such that
$\mbox{curl }{\bf g}=0$ in this language corresponds to a pure gauge
transformation:
\begin{equation}
\gamma(t) = \exp \left\{ i \int d^{3}{\bf x} \; g_{i}({\bf x}) \,
\Pi_{i}({\bf x}, t)\right\}
\end{equation}
By gauge-invariance
$\langle 0 | \gamma(t) | 0 \rangle = 1$.
Performing the Wick rotation to Euclidean space, we obtain:
\begin{equation}
\mu_{E}({\bf y}, y_4) = \exp\left\{- g\;\int d^{3}{\bf x} \; b_{i}({\bf x},
{\bf y}) \, \Pi_{i}({\bf x}, y_4)\right\}\end{equation} From
now on, we will be interested only in the Euclidean quantity and
drop for simplicity the $E$ subscript.
We will first compute $\langle \mu\rangle\equiv
\langle 0|\mu_E({\bf y},y_4)|0\rangle$ for free photons.
Rescaling the fields by a factor $1/\sqrt{\beta}$
with $\beta = 1/e^2$, we have
\begin{eqnarray}
\langle \mu \rangle &=& \frac{1}{Z}{\displaystyle \int} {\cal D}A
\left\{
\exp \left[
-\frac{\beta}{4} {\displaystyle \int}
d^4 x F_{\mu\nu} F_{\mu\nu} \right]\right.\\
&&\left.\times\exp \left[-\beta {\displaystyle \int} d^3 x
 F_{0i}({\bf x},y_4) b_{i}({\bf x},{\bf y}) \right]\right\} \nonumber
\end{eqnarray}
The integral is Gaussian and can be directly computed giving:
\[
\langle \mu \rangle = \exp \left\{ \frac{\beta}{2}
\int \frac{d^{4}k}{(2\pi)^4} \langle F_{0i}(k)
F_{0j}(-k) \rangle b_{i}(-{\bf k}) \; b_{j}({\bf k}) \right\}
\]
\begin{equation}
\langle F_{0i}(k) F_{0j}(-k)\rangle =
\delta_{ij} -
\frac{\displaystyle {\bf k}^2\delta_{ij} - k_i k_j}
{\displaystyle k_0^2 + {\bf k}^2}\label{fij}
\end{equation}
If ${\bf b}$ is such that $k_i b_i(k) = 0$, we have, performing the integral
over $k_0$ in the second term,
\begin{equation}
\langle \mu \rangle = \exp \frac{\beta}{2}\left\{\int
\frac{d^{4}k}{(2\pi)^4} |{\bf b}(k)|^2 -
\frac{1}{2} \int
\frac{d^{3}k}{(2\pi)^3} |{\bf k}| \, |{\bf b}(k)|^2 \right\}
\label{defmu}
\end{equation}
In the analogous calculation
for a  gauge transformation $\gamma$, $g_i({\bf k})\propto k_i$, the second
term of eq.(\ref{fij}) does not contribute, and
\begin{equation}
\langle \gamma \rangle = \exp \left\{ \frac{\beta}{2} \int
\frac{d^{4}k}{(2\pi)^4} |{\bf g}(k)|^2 \right\}
\end{equation}
We realize  that the first term in the exponent
of eq.(\ref{defmu})
is a normalization
which can be subtracted by taking instead of $\langle \mu \rangle$ the
ratio $
\langle \bar\mu \rangle = \langle \mu \rangle/\langle \gamma\rangle $
where $\langle \gamma \rangle$ is defined by means of any gauge
transformation ${\bf g}$, such that:
\begin{equation}
\int \frac{d^{4}k}{(2\pi)^4} |{\bf g}(k)|^2 =
\int \frac{d^{4}k}{(2\pi)^4} |{\bf b}(k)|^2
\label{eq:norm}
\end{equation}
Then for free photons, which,
in the compactified version of the theory,
correspond to the deconfined phase, $\beta \gg \beta_c$
\begin{equation}
\langle \bar\mu \rangle = \exp \left\{
-\frac{\beta}{4} \int
\frac{d^{3}k}{(2\pi)^3} |{\bf k}| \, |{\bf b}(k)|^2 \right\}
\label{eq:pert}
\end{equation}
The integral in the exponent, once regularized at small distances,
tends to $+\infty$ as $V\to\infty$.
In the infinite volume limit $\langle \bar\mu \rangle=0$ as it should
be, since the perturbative vacuum has zero magnetic charge. The same
situation appears when computing the overlap of the Fock vacuum to the
Bogolubov rotated vacuum in a superconductor~\cite{r23}.

An alternative way of looking at the normalization factor in
eq.(\ref{defmu}) is to go back to the very definition of Feynman path
integral: when computing the vacuum expectation value of an operator like $e^{i
p(t) a}$, after the usual discretization of time is performed in intervals of
size $\delta$, the operator appears in a matrix element of the form
\begin{eqnarray*}
\langle x_{n+1}|e^{-i H \delta}\,e^{i p a}\,|x_n\rangle
&=&\int\frac{\displaystyle d p}{\displaystyle(2\pi)}
\langle x_{n+1}|e^{-i H \delta}\,e^{i p a}\,|p\rangle\langle p||x_n\rangle\\
&=&\int\frac{\displaystyle d p}{\displaystyle(2\pi)}
\,e^{-i(\frac{p^2}{2m} + V(x_n))}
e^{i p a} e^{i p (x_{n+1}-x_n)}
\end{eqnarray*}
The integral over $p$ can be performed giving
\[e^{-i(x_{n+1}-x_n)^2\frac{m}{2}+i a(x_{n+1}-x_n)m+V(x_n)}
e^{-i a^2}\]
The first factor is the lagrangian definition, the second factor corresponds to
the subtraction operated in going from $\langle\mu\rangle$ to
$\langle\bar\mu\rangle$ in eq.(\ref{eq:pert}).
\section{Lattice formulation. $U(1)$ gauge theory.}
A lattice version of the operator $\mu$ is obtained by  replacing
$e F_{0i}$ by the plaquette $P_{0i}$, or better, by its imaginary part:
$e F_{0i} \to Im\,P_{0i}$
and discretizing the field $b_i$.
Then the disorder parameter becomes:
\[
\langle \mu \rangle =
\frac{1}{Z}
{\displaystyle \int} {\cal D}U
\exp\{-\beta [ S
+ \sum_{n_0=y_4} b_{i}({\bf n},{\bf y}) Im\,P_{0i}(n)] \}
\]
or, if we want to cancel the unwanted normalization, we can divide by:
\[
\langle \gamma \rangle =
\frac{1}{Z}
{\displaystyle \int} {\cal D}U \exp\{-\beta [ S
+ \sum_{n_0=y_4} g_{i}({\bf n}) Im\,P_{0i}(n)]\}
\]
obtaining
\begin{equation}
\langle \bar\mu \rangle = \frac
{{\displaystyle \int}\! {\cal D}U
\exp\left\{-\beta [ S
+ {\displaystyle\sum_{n=y_4}} b_{i}({\bf n},{\bf y}) Im\,P_{0i}(n)
]\right\}} {{\displaystyle \int} {\cal D}U
\exp\left\{-\beta [ S
+ {\displaystyle\sum_{n=y_4}} g_{i}({\bf n}) Im\,P_{0i}(n) ]\right\}}
\end{equation}
We stress once again  that any gauge function $g_{i}$ is acceptable,
provided the normalization condition (\ref{eq:norm}) is satisfied.
In the following we shall use Wilson's action
$S = \frac{1}{2}\sum_{n,\mu,\nu}(1-P_{\mu\nu})$.

If we blindly compute $\langle \mu \rangle$ or  $\langle \bar\mu
\rangle$ by numerical simulations, a first technical difficulty
arises. We are faced with the usual problems encountered in computing
quantities like a partition function, which are exponentials of
extensive quantities, proportional to the number of degrees of
freedom. The distribution of the values is not Gaussian and the error
does not decrease by increasing statistics (see, e.g.~\cite{HHN},
where the same problem appears in a different context). To avoid that,
we will compute the quantity:
\begin{equation}
\rho = \frac{d}{d\beta} \log \langle \bar \mu \rangle =
\frac{d}{d\beta} \log \langle \mu \rangle -
\frac{d}{d\beta} \log \langle \gamma \rangle
\label{defrho}
\end{equation}
At $\beta=0$, $\langle \mu \rangle = \langle \gamma \rangle = 1$, and
therefore:
\begin{equation}
\langle \bar \mu \rangle = {\rm exp}
\left[\int_{0}^{\beta} d\beta^{\prime}
\rho(\beta^{\prime}) \right]
\end{equation}
Putting
$S_b = \left.\sum b_{i}(n) Im P_{0i}(n)\right|_{n_0=0}$ and
$S_g =$ $\left.\sum g_{i}(n)\right.$ $\left. Im
P_{0i}\right.$ $\left.\right|_{n_0=0}$
we get:
\begin{equation}
\rho = \langle S + S_g \rangle_{S+S_g} -
\langle S + S_b \rangle_{S+S_b}
\end{equation}
which can be evaluated by numerical simulations. The subscript on the average
indicates the action defining the Feynman integral.
The two quantities on the {\em rhs} have the same strong coupling
expansion. Thus, the use of $\langle \bar\mu \rangle$ instead of $\langle
\mu \rangle$, besides producing a cancellation of the spurious
normalization of the Feynman path integral, can help in eliminating
the lattice artefacts produced by the discretization which can spoil the
continuum limit. This brings us to a second, more physical difficulty,
which is the continuum limit. In fact, while for {\it QCD\/}, which is
asymptotically free, we expect that, at sufficiently high $\beta$,
lattice artefacts should cancel, for a model like $U(1)$ this point is
not so clear in principle. In Ref.~\cite{FM} a proof is given of
monopole condensation in the confined phase of $U(1)$, defining a
disorder variable for the Villain action. We have checked that our
$\langle \mu \rangle$ operator exactly coincides with the one of
Ref.~\cite{FM}, when we use the Villain action: we expect that for
Wilson action the same will hold. We have then computed numerically
$\rho$.

For a {\em good} order parameter, we would expect $\rho$ to be zero,
or $\langle\bar\mu\rangle = 1$
below
the critical value of the coupling $\beta_c$ and then to show a large
negative peak around $\beta_c$, corresponding to a drop to zero of
$\langle \bar\mu \rangle$. At larger values of $\beta$, we have free photons
and Eq.~(\ref{eq:pert}) should hold.

Figure~1 shows the behaviour of $\rho$ for a $12^4$ lattice,
for a monopole in the center of the space lattice. In order
to be able to identify the signal as a genuine physical result (i.e.
not due to lattice artefacts), we have performed a number of checks:
\par\noindent{\it 1.}
We have changed the form of $b_i$ by a gauge transformation to
get the Wu-Yang expression of the monopole potential. The result does
not change qualitatively.
\par\noindent{\it 2.}  $\langle \gamma \rangle$ shows
practically no signal at $\beta_c$ within the errors, and does not change
appreciably by changes of $g_i$.
To test that we have computed $\langle \gamma\rangle$ for two different choices
of $g_i({\bf x})$,  both satisfying the condition Eq.(\ref{eq:norm}):
$\langle \gamma_1\rangle$ for ${\bf g}_1({\bf x}) = {\rm const.}$ and
$\langle \gamma_2\rangle$ for $g_2({\bf x})\propto {\bf x}/|{\bf x}|^2$.
In Fig.2 we display
$ \rho_{gauge} = \frac{d}{d\beta}\ln\frac{\langle \gamma_2\rangle}
{\langle \gamma_1\rangle}$
and $\rho = \frac{d}{d\beta}\ln\frac{\langle \mu\rangle}
{\langle \gamma_1\rangle}$ for a $6^4$ lattice.
$\rho_{gauge}$ shows no relevant signal at $\beta_c$.
\par\noindent{\it 3.}
 We have measured the correlation function between a
monopole antimonopole pair at large time distance, in the same position in
space.
To do that we define
$ C(d) = \langle\mu({\bf 0},0)\,\mu({\bf 0},d)\rangle$ and
\[ S_{b\bar b}(d) =
\sum_{{\bf n}} b_i({\bf n},{\bf 0}) \,
[ Im P_{0i}({\bf n},0)
-  Im P_{0i}({\bf n},d) ] \]
We measure $\frac{d}{d\beta}\ln\frac{C(d)}{\langle\gamma\rangle^2} =
\rho_{(b,\bar b)}(d)$
or
\[ \rho_{(b,\bar b)}(d) =
2 \langle S + S_g\rangle_{S+S_g} - \langle S + S_{b\bar b}(d)\rangle_{
S + S_{b\bar b}} - \langle S\rangle_S\]
Since $\left. \frac{C(d)}{\langle\gamma\rangle^2}\right|_{\beta=0} = 1$
we have
\begin{equation}\frac{C(d)}{\langle\gamma\rangle^2}
 = {\rm exp}\left[\int_0^\beta \rho_{(b,\bar b)}(d) d\beta
\right]
\end{equation}
By the cluster property we should have at large $d$ that
$C(d) \to \langle \mu({\bf 0},0)\rangle^2$, or
$\rho_{(b,\bar b)}(d) \to 2\rho$. Fig.3 shows that this expectation is indeed
verified.
We notice that the height of the negative peak of $\rho$ at $\beta_c$ increases
with volume [see Fig.1 and Fig.2]. The value of $\beta_c$ as defined by the
position of our peak is $\beta_c = 1.01(1)$ for a $6^4$ lattice and $\beta_c =
1.009(1)$ for a $12^4$ lattice.

We have taken monopole charge $n=4$ to get a good signal
with a relatively low statistics (tipically $10^4$ configurations per value of
$\beta$): smaller charges ($n=1,2$) give similar results but the signals are
smaller and more noisy. Our statistical errors are shown in the figures, when
they are larger than the symbols used.
\section{$SU(2)$ gauge theory. Monopole condensation and confinement}
The monopoles which should condense in $QCD$ vacuum in the confined phase are
Dirac monopoles of the $U(1)$ field defined by the abelian
projection\cite{tHO}.
The density of current of such monopoles is
\[ j_\nu = \partial^\mu F^*_{\mu\nu}\]
where, in the notation of ref.\cite{tHO2}
\begin{eqnarray}
&&F_{\mu\nu} =
\frac{\displaystyle1}{\displaystyle|\vec Q|}Q_a G^a_{\mu\nu}
-
\frac{\displaystyle1}{\displaystyle g|\vec Q|^3}
\varepsilon_{abc}Q^a(D_\mu Q^b)(D_\nu Q^c)\label{eq:nonabeliana}\\
&&\quad a,b,c = 1,2,3\nonumber
\end{eqnarray}
$Q^a$ is the Higgs field in the Georgi - Glashow model. Abelian projection is a
gauge transformation which brings $Q$ in the third direction, $Q = (0,0,Q^3)$.

In $QCD$ the role of $Q^a$ should be played by some operator
trasforming in the adjoint representation\cite{tHO}: what this operator is, if
any, is the problem under investigation.

After abelian projection the second term of eq.(\ref{eq:nonabeliana})
gives zero and $F_{\mu\nu} = G^3_{\mu\nu}$.

As a first candidate for $Q^a$ we have chosen $p^a = \frac{1}{2}{\rm Tr}(-i \ln
P\,\sigma^a)$ where $P$ is the Polyakov line\cite{tHO}, $exp(\int A_o dx^0)$.
This choice corresponds to take $A_0$ as the Higgs field.
For
that choice $D_0 Q = 0$ and
\[ F_{0i} = \frac{\displaystyle p^a}{\displaystyle|\vec p|} G^a_{0i}\]
Since the expression (\ref{eq:nonabeliana}) is gauge invariant, we do not
need to
perform any gauge transformation to make $\ln P$ diagonal.
We simply define the gauge invariant quantity
\[ F_{0i}= \frac{1}{2}\frac{\displaystyle {\rm Tr}(-i\,\ln P\, G_{0i})}
{\displaystyle|\vec p|}\]
The operator creating the relevant monopole is
\begin{equation}
\mu({\bf y},t) = {\rm exp}\{-\int d^3{\bf x} b_i({\bf x},{\bf y}) F_{0i}({\bf
x},t)\} \end{equation}
Exactly as for $U(1)$ we can define
$\langle\bar\mu\rangle = \langle\mu\rangle/\langle\gamma\rangle$
where
\[\langle\gamma(t)\rangle =
{\rm exp}\{-\int d^3{\bf x} g_i({\bf x}) F_{0i}({\bf
x},t)\}\]
We shall take for $g_i({\bf x})$ a constant field, such that
eq.(\ref{eq:norm}) is satisfied.

The lattice version of $F_{0i}$ will be
\[ F_{0i} = \frac{1}{2}\frac{\displaystyle {\rm Tr}\{-i\,\ln P\,\Pi_{0i}\}}
{|\vec p|}\]
and
\begin{equation}
\langle \bar\mu \rangle = \frac
{{\displaystyle \int}\! {\cal D}U
\exp\left\{-\beta [ S
+ {\displaystyle\sum_{n=y_4}} b_{i}({\bf n},{\bf y}) F_{0i}(n)
]\right\}} {{\displaystyle \int} {\cal D}U
\exp\left\{-\beta [ S
+ {\displaystyle\sum_{n=y_4}} g_{i}({\bf n})  F_{0i}(n) ]\right\}}
\end{equation}
We will define $\rho$ as in eq.(\ref{defrho}), and therefore again:
\[
\langle \bar \mu \rangle = {\rm exp}
\left[\int_{0}^{\beta} d\beta^{\prime}
\rho(\beta^{\prime}) \right]
\]
As in $U(1)$ we measure the behaviour of $\rho(\beta)$.
across the deconfining phase transition
The simulations are performed on asymmetric lattices, with
temporal size $N_T$ smaller then spacial size $N_S$.

Fig. 4,5,6,7 show the behaviour of $\rho$ for $8^3\times4$, $12^3\times4$,
$12^3\times6$ and $16^3\times6$ lattices. The behaviour of $ \rho$
is similar to the $U(1)$ case: $\rho$ is compatible with zero below the
deconfining transition, has a sharp negative peak at the deconfining phase
transition, and a negative\hfill
 tail at high $\beta$, corresponding to a phase of free
``photons'' (see eq.(\ref{eq:pert}). Qualitatively the shape of
$\langle\bar\mu\rangle$ is \[ \langle\bar\mu\rangle = \theta(\beta_c - \beta)\]
The negative peak increases with space volume at fixed $N_T$ (compare
$8^3\times4$ with $12^3\times4$, $12^3\times6$ with $16^3\times6$) which means
that as $V\to \infty$, for $\beta > \beta_c$, $\langle\bar\mu\rangle\to 0$.

The peak is placed exactly at the official location of the phase transition
for $N_T=4$\cite{Engels}
and
moves as expected from renormalization group from $N_T=4$ to $N_T=6$.

The usual order parameter $|\langle P\rangle|$ is plotted for comparison in the
figures.

The depth of the peak does not change appreciably from $8^3\times4$ to
$12^3\times6$, showing that the signal mainly depends on the physical volume of
the lattice, i.e. on its size in ${\rm fm}^{3}$, and not on the geometrical
volume of it.

We have also measured the correlation function
\[\langle\bar\mu(10,0,0,0),\bar\mu(0,0,0,0)\rangle\]
on a $8^2\times20\times4$ lattice to check the cluster property, and tested
that
it is equal to the square of the disorder parameter on the same lattice
(Fig. 8).

We can finally conclude that the monopoles defined by abelian projection
identified by the Polyakov loop are charges of an $U(1)$ symmetry which is
spontaneously broken in the confined phase, and restored above $\beta_c$, or
that confinement is produced by dual superconductivity.
\section{Concluding remarks}
We have demonstrated that confinement in $SU(2)$ gauge theory is due to dual
superconductivity of the vacuum. We have identified the monopole charges which
condense and produce it. Our method consists in a direct detection of
spontaneously breaking of the $U(1)$ symmetry corresponding to monopole charge.

The method has been successfully checked on $U(1)$ compact gauge theory.

A number of open questions are currently under investigation
\begin{itemize}
\item[i)] Are there different abelian projections defining monopoles which
condense in $QCD$ vacuum? Is t'Hooft's idea about the
physical equivalence of different abelian projections correct?
In particular we are investigating the so called maximal abelian
gauge, defined by the gauge transformation which maximises the quantity
\[\sum_{n,\mu}{\rm Tr}\left\{
U_\mu(n) \sigma_3 U^\dagger_\mu(n) \sigma_3\right\}
\]
On this gauge there is a lot of evidence collected in favour
of dual superconductivity\cite{Schier,r16}.
\item[ii)] The extension to $SU(3)$ of the above construction.
\item[iii)] The application of our method to other systems, like e.g. the $3d$
$x$-$y$ model, in which a phase transition is driven by the condensation of
solitons.
\end{itemize}

%
\vfill\eject{\centerline{\bf \large FIGURE CAPTIONS}}
\vskip0.2in
\vskip0.2in\par\noindent Fig.1\,\,{$\rho$ versus $\beta$ on a $12^4$ lattice
\label{fig1}}
\vskip0.2in\par\noindent Fig.2\,\,{$\rho$ and $\rho_{gauge}$ versus $\beta$
on a $6^4$ lattice.
\label{fig2}}
\vskip0.2in\par\noindent Fig.3\,\,{$\rho_{b\bar b}(d)$ versus $\beta$ at
$d=4,7,9$, compared to
$2\rho$.
\label{fig3}}
\vskip0.2in\par\noindent Fig.4\,\,{$\rho$ (circles) and $\langle
| P|\rangle$ (squares), $8^3\times4$ $SU(2)$ gauge
theory.
\label{fig4}}
\vskip0.2in\par\noindent Fig.5\,\,{$\rho$ (diamonds) and $\langle|
P|\rangle$ (squares),
$12^3\times4$ $SU(2)$ gauge theory.
\label{fig5}}
\vskip0.2in\par\noindent Fig.6\,\,{
$\rho$ (triangles) and $\langle| P|\rangle$ (squares), $12^3\times6$
$SU(2)$ gauge
theory.
\label{fig6}}
\vskip0.2in\par\noindent Fig.7\,\,{$\rho$ (open squares) and $\langle|
P|\rangle$ (squares),
$16^3\times6$ $SU(2)$ gauge theory.
\label{fig7}}
\vskip0.2in\par\noindent Fig.8\,\,{$\rho_{b\bar b}(d)$ at $d=10$ (circles)
compared to $2\rho$ (triangles) and their difference (squares).
$SU(2)$ gauge
theory.
\label{fig8}}
\end{document}